\documentclass{sig-alternate-10pt}

\usepackage[usenames,dvipsnames,svgnames,table]{xcolor}
\usepackage{textcomp}
\usepackage{url}

\usepackage{listings}

\definecolor{listinggray}{gray}{0.95}
\definecolor{lbcolor}{rgb}{0.95,0.95,0.95}
\definecolor{Darkgreen}{rgb}{0, 0.3, 0.0}

\usepackage[htt]{hyphenat}
\newcommand\code[1]{{\texttt {#1}}}

\begin{document}



\conferenceinfo{ICN '16}{Sept. 26-28 Kyoto, Japan}

\title{A new NS3 Implementation of CCNx 1.0 Protocol}

\numberofauthors{1}

\author{
\alignauthor
Marc Mosko, Ramesh Ayyagari, Priti Goel, Eric Holmberg, Mark Konezny\\
       \affaddr{Palo Alto Research Center}\\
       \affaddr{3333 Coyote Hill Road}\\
       \affaddr{Palo Alto, CA, 94304 USA}\\
       \email{Corresponding Author: marc.mosko@parc.com}
}

\maketitle
\begin{abstract}
The \code{ccns3\-Sim} project is an open source implementation of the CCNx 1.0 protocols
for the NS3~\cite{ns3} simulator.  We describe the implementation and several important features
including modularity and process delay simulation.
The \code{ccns3\-Sim} implementation is a fresh NS3-specific implementation. Like NS3 itself, it uses C++98 standard, NS3 code style, NS3 smart pointers, NS3 xUnit, and integrates with the NS3 documentation and manual. A user or developer does not need to learn two systems.
If one knows NS3, one should be able to get started with the CCNx code right away. 
A developer can easily use their own implementation of the layer 3 protocol, layer 4 protocol, forwarder, routing protocol,
Pending Interest Table (PIT) or Forwarding Information Base (FIB) or Content Store (CS).  A user may configure
or specify a new implementation for any of these features at runtime in the simulation script.
In this paper, we describe the software architecture and give examples of using the simulator.
We evaluate the implementation with several example experiments on ICN caching.
\end{abstract}



\section{Introduction}
The \code{ccns3Sim}~\cite{ccns3Sim} open source project 
is an NS3~\cite{ns3} implementation of the CCNx 1.0 protocol specifications \cite{I-D.irtf-icnrg-ccnxsemantics, I-D.irtf-icnrg-ccnxmessages}.
It is released as a source code module for the NS3 simulator.  It runs with an unmodified NS3 simulator,
though one patch is required to add an Ethertype to the PPP implementation.  \code{ccns3Sim} includes many examples and documentation.

The \code{ccns3Sim} implementation is a fresh NS3-specific implementation. Like NS3 itself, it uses C++98 standard, NS3 code style, NS3 smart pointers, NS3 xUnit, and integrates with the NS3 documentation and manual. A user or developer does not need to learn two systems.
If one knows NS3, one should be able to get started with the CCNx code right away. 
A developer can easily use their own implementation of the layer 3 protocol, layer 4 protocol, forwarder, routing protocol,
Pending Interest Table (PIT) or Forwarding Information Base (FIB) or Content Store (CS).  A user may configure
or specify a new implementation for any of these features at runtime in the simulation script.

The general architectural model of \code{ccns3Sim} is to use an abstract base class as an \emph{interface}
and then provide an implementation class of that interface.  For example, the \code{CCNxPit} abstract base class
defines the API of the PIT table and the implementation class \code{CCNxStandardPit} realizes a PIT
based on the current specifications.  Because the abstract class inherits from \code{ns3::Object}, one
can use run-time binding, making it easy for a user to substitute new implementations.

We introduce the Name Flooding Protocol (NFP), a simple distance-vector routing protocol for CCNx.
It provides basic loop-free equal-cost multipath routing for CCNx names.  Each \emph{anchor} for
a prefix (device advertising a prefix) uses its own sequence number for each advertisement, which
are then ordered by distance.  Each anchor is an independent successor graph for the prefix
and will induce its own equal-cost multipath graph.  If two or more anchors advertise the same prefix,
NFP allows unequal cost multipath to each anchor.

The ndnSim project~\cite{mastorakis2015ndnsim} is a customized NS3 distribution that uses the regular NDN open
source code ndn-cxx (NDN C++ library with eXperimental eXtensions) and nfd (NDN Forwarding Daemon).  This approach
has a significant advantage in that it uses the same libraries as regular NDN applications and forwarders, so there is
high code re-use and the simulator stays up-to-date with the non-simulation release.  The disadvantage of this
approach is that simulation must use the production code, it introduces two object models, two memory management models,
and two coding style models.  It may also make certain instrumentation that one would like to see in simulation
difficult, because the ndn-cxx and nfd libraries were not designed with simulation in mind.  For example,
it can be difficult to simulate processing delays in data structures.

ccnSim~\cite{6654874} is a C++ simulator implemented in Omnet++.  The authors claim it is a highly scalable
simulator able to support content stores with up to $10^6$ chunks and catalogs of
up to $10^8$ files.  The current version, ccnsim-v0.4, may support much larger catalogs
and operates up to 100x faster than the previous version.  ccnSim supports several
caching mechanisms (decision strategies), such as leave copy everywhere, leave copy down,
and betweeness centrality.  It also supports several cost and replacement functions to determine if a content
is cached what what gets evicted.  Our present work, \code{ccns3Sim}, is a more general
purpose simulator for transport protocols, routing protocols, and other
protoocol features, it is not a highly optimized for content caching simulations.
NS3's operational model of serializing and deserializing each packet at each hop means that content store
copies are not memory references to a common repository.  This makes the
initial code release of \code{ccns3Sim} inefficient for large content caching experiments compared to ccnSim.

The remainder of the paper is organized as follows.  Section~\ref{sec:architecture} describes \code{ccns3Sim}'s
software architecture. 
It also describes the robust delay model and shows how to implement both
input delays and processing delays.
Section~\ref{sec:layer3} describes the layer 3 components, and in particular the forwarder and associated tables.  
Section~\ref{sec:layer4} describes the layer 4 components, which we call the \code{CCNxPortal} interface.
Section~\ref{sec:routing} describes the routing protocol and the details of NFP.  
Section~\ref{sec:apps} describes the application model and the example consumer-pro\-duc\-er app.
Section~\ref{sec:experiments} presents several experiments conducted with \code{ccns3Sim}.
Section~\ref{sec:conclusion} concludes the paper and summarizes it's contributions.


\section{Node Architecture}\label{sec:architecture}
\code{ccns3Sim} is designed such that a single \code{ns3::Node} may have zero or more
CCNx applications running on it, plus a forwarder and routing protocol.  An application
running on a node uses one or more \code{CCNxPortal} interfaces to pass messages
to the local forwarder.  Each application instantiates its own Portals, similar to how
an Internet application uses Sockets.  A routing protocol is similar to an application
in that it uses one or more Portals to communicate with peers, but it derives
from \code{CCNxRoutingProtocol} instead of \code{CCNxApplication}.  A routing
protocol also uses the \code{CCNxForwarder} interface on a node to
manipulate the FIB inside the forwarder.
\code{ccns3Sim} can operate with any NS3 \code{NetDevice}.  We have tested
with point-to-point, CSMA, WiFi, and virtual devices using UDP tunnels.

\begin{figure}
\centering
\includegraphics[width=.95\linewidth, trim=.2in .2in .2in .2in, clip]{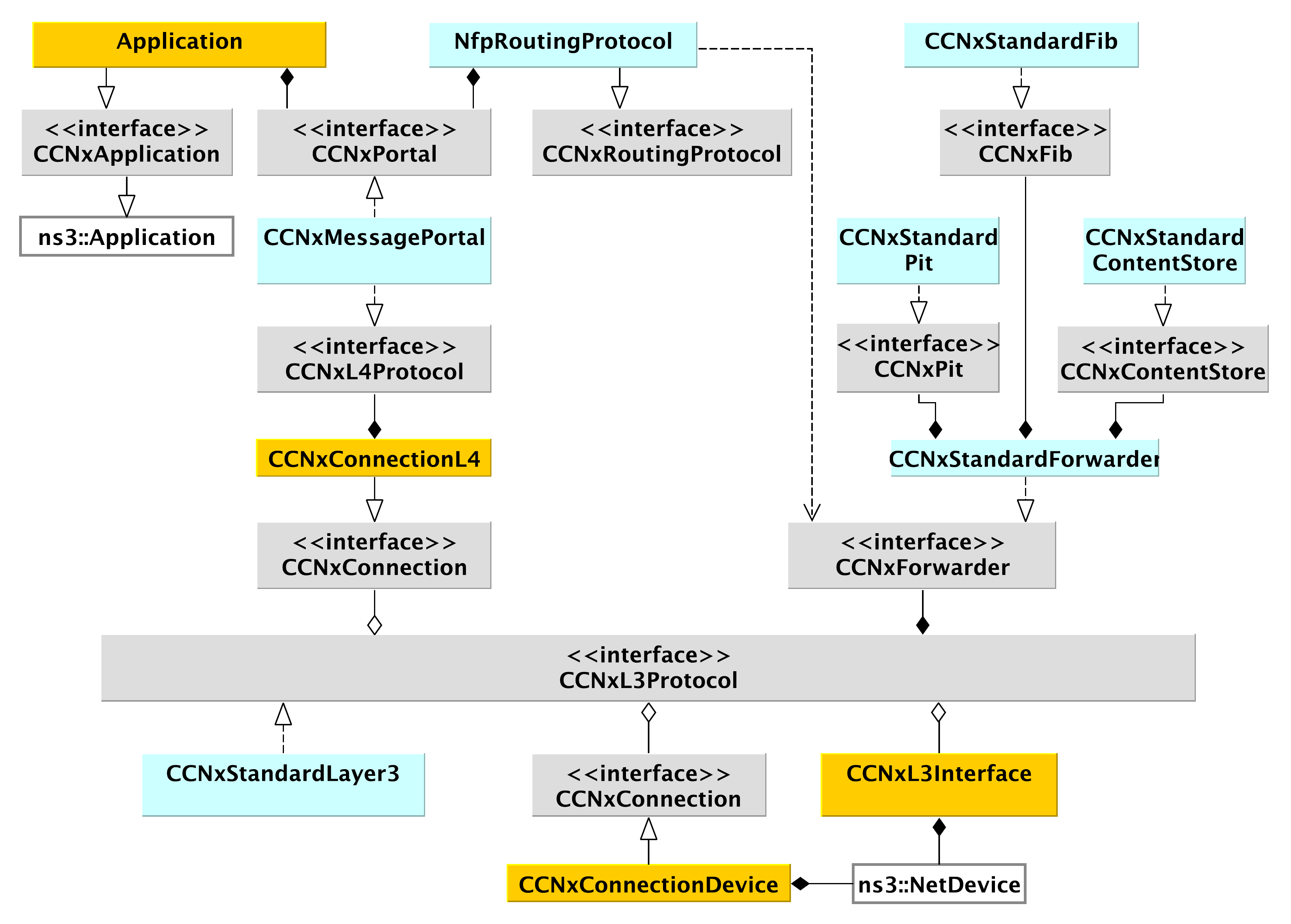}
\caption{CCNx Class Architecture}
\label{fig:architecture}
\end{figure}

The CCNx NS3 architecture is shown in Fig.~\ref{fig:architecture}.  Every major component is described
by an interface (abstract base class) and inherits from the \code{ns3::Object}.  This allows run-time
late binding of each piece to a user-specified implementation in the simulation script.
The core of the architecture is the \code{CCNxL3Protocol}, which is the switching fabric for
the layer 3 protocol.  It is connected to network devices below and layer 4 protocols above.
There is also a special interface \code{CCNxForwarder} to the forwarding engine, which is
responsible for selecting the egress connection of each input packet.   We describe each of
these components in more detail below.

The CCNx node architecture differs from the NS3 Internet model in that we have
separated the routing protocol from the forwarding process.  In the NS3 model, the
routing protocol, in addition to exchanging routing messages with peers and executing
a routing process, also handles calls to \code{RouteInput()} and \code{RouteOutput()}.
In the CCNx architecture, the routing protocol only exchanges messages with
peers and executes a routing process.  It maintains a Routing Information Base (RIB).
Using the \code{CCNxForwarder} interface, the routing protocol calls
\code{AddRoute()} and \code{RemoveRoute()} to manage the Forwarding
Information Base (FIB) inside the forwarder.  The forwarder, via the
\code{CCNxForwarder} interface, handles calls to \code{RouteInput()} and \code{RouteOutput()}.

\begin{figure}
\centering
\includegraphics[width=.95\linewidth, page=9, trim=.6in .8in .1in .2in, clip]{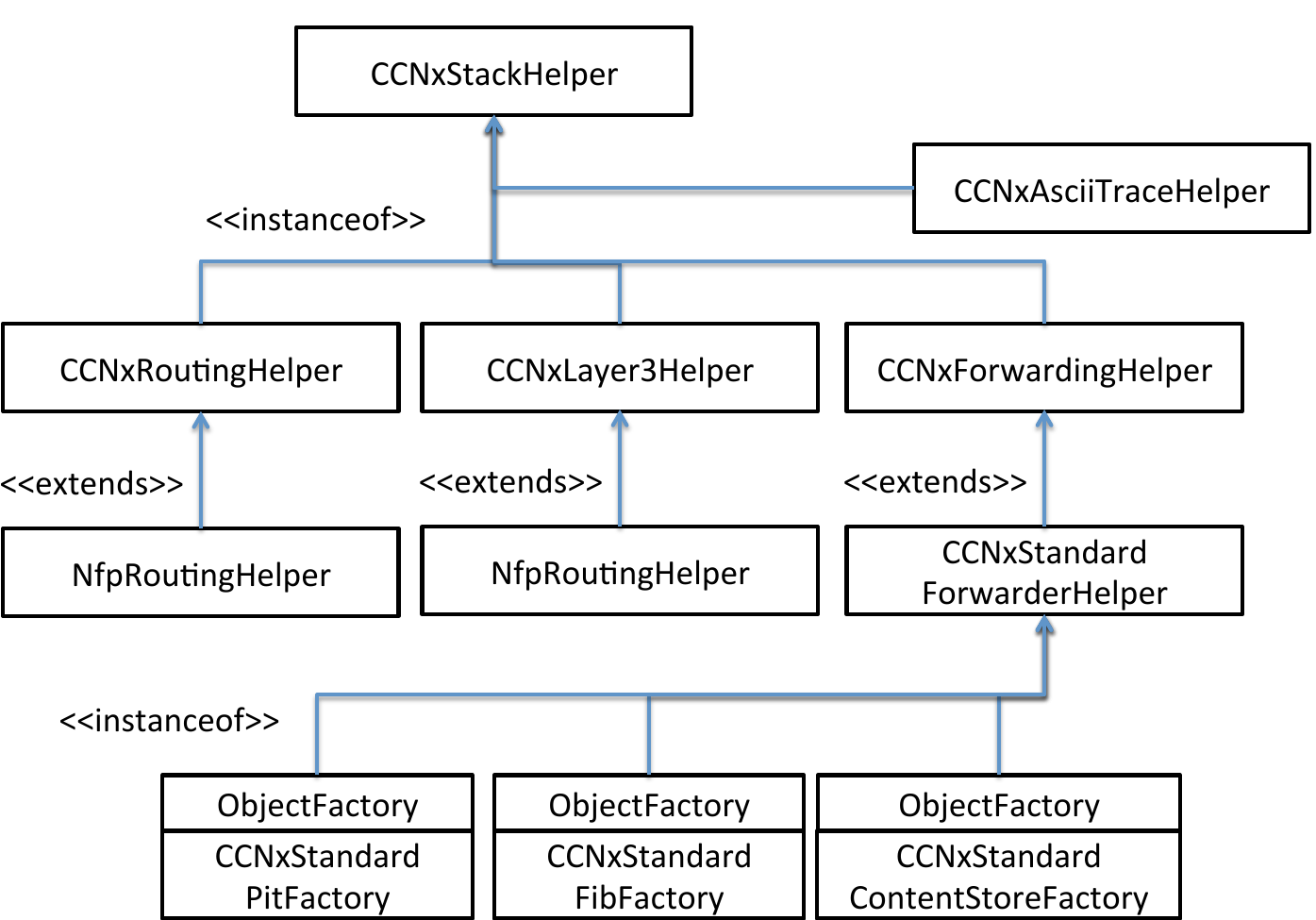}
\caption{CCNxStackHelper}
\label{fig:helpers}
\end{figure}

The user selects each component via the simulation script.  Figure~\ref{fig:helpers}
shows the set of ``helpers'' for use in the simulation script.  The top-most helper
is the \code{CCNx\-Stack\-Helper}.  It is responsible for the overall installation of
CCNx on an \code{ns3::Node} or \code{ns3::Node\-Con\-tainer}.
Each major architectural component -- the layer 3 implementation,
the forwarder implementation, and the routing protocol -- has its own
helper.  

In NS3 terminology, a Helper is a factory that can create
an object and then \emph{aggregates} it to an NS3 node.  Aggregating
an object to a node makes an association between the class and the
node, so at a later time one can ask a node for a specific implementation.
This is, for example, how the \code{NfpRoutingProtocol} uses the
\code{CCNxForwarder} interface without needing to know exactly
which implementation is being used.  When the forwarder implementation,
such as \code{CCNxStandardForwarder} aggregates on to a node,
it can be retrieved via its base class.

\begin{itemize}
\item \emph{NfpRoutingHelper}: Used to configure and install the NFP routing protocol.
\item \emph{CCNxStandardLayer3Helper}: Used to configure and install the
standard Layer3.  This component manages the set of Layer 2 interfaces,
the Layer 3 peers and Layer 4 protocols.
\item \emph{CCNxStandardForwarderHelper}: The for\-ward\-er hel\-per installs
the RFC-draft compliant forwarder, which its associated PIT, FIB, and Content Store.
Note that the table factories are specific to the forwarder helper implementation.
A label-swapping forwarder, for example, that does not have a PIT table does
not need to expose an \code{ObjectFactory} for creating a PIT.
\item \emph{CCNxStandardPitFactory}: The factory from which PIT tables are created
with a given set of parameters and implementation.
\item \emph{CCNxStandardFibFactory}: The factory from which FIB tables are created
with a given set of parameters and implementation.
\item \emph{CCNxStandardContentStoreFactory}: The factory from which Content Stores are created
with a given set of parameters and implementation.
\end{itemize}

The code snippit in Listing~\ref{stackhelper} shows a portion of a simulation script
that uses the \code{CCNxStackHelper} to install the CCNx protocol on all the
nodes in a \code{NodeContainer}.

\begin{lstlisting}[caption={Installing CCNx on NodeContainer},label=stackhelper]
NodeContainer nodes;
nodes.Create(5);

CCNxStackHelper ccnx;
ccnx.Install (nodes);
\end{lstlisting}

The CCNxPortal abstraction is the layer 4 protocol from the applications's point of view. It has functions to Send(), SendTo(), Recv(), and RecvFrom() using a \code{CCNxMessage}. We provide one implementation class, \code{CCNxMessagePortal}, which is a simple 1-for-1 protocol without any transport features (like a UDP). The CCNxMessage class is the base for \code{CCNxInterest} and \code{CCNxContentObject}. The class \code{CCNxL4Protocol} is the layer 4 abstraction from the layer 3 point of view. An implementation class like \code{CCNxMessagePortal} inherits from both \code{CCNxPortal} and \code{CCNxL4Protocol}.

One can run CCNxL3Protocol over any Layer 2 \code{NetDevice}s. We have used it over point-to-point, CSMA, WiFi, and \code{VirtualNetDevice} in our unit tests.  One could use CCNx over IP tunnels (e.g. UDP tunnels) using a \code{VirtualNetDevice} wrapping a UDP socket.

\code{ccns3Sim}'s packet coding is wire-format compatible with the standard CCNx 1.0
TLV format.  We have, for example, taken PCAP traces from NS3 and replayed them on
real networks to our software forwarders.  The current NS3 implementation is only
partially configurable for modular codecs, but we expect to have fully configurable
codecs in the near future.  Because CCNx 1.0 uses the hierarchical nesting of
TLV types to determine containment, we will use a notation like SNMP MIB OIDs
to specify codecs.  For example, TLV type 1 is Interest message and TLV type 0 is
it's CCNxName, so ``.1.0'' would specify the codec for the name.  TLV type 1 is
the Payload of the Interest, so ``.1.1'' is used to lookup the payload's codec, and so on.
This model allows a user to either replace a specific codec or introduce new TLV types and
new codecs at run time.  A codec may be used by multiple OIDs.  For example ``.1.0'' and
``.2.0'' are both CCNxName codecs.  The first is for an Interest message and the second
is for a Content Object message.

\subsection{Modeling delays}\label{sec:delays}

The \code{CCNxDelayQueue} is a general queuing class to delay events.  Several of
the major CCNx forwarding components have delay queues to model computation time.
We use the terminology \emph{input delay} to mean a delay that only depends on
the input to the delay queue, not the processing done.  The term \emph{processing delay}
means a delay that depends on the processing, such as the number of tables consulted
or the number of name components searched.  Both use the same data structure,
it is only a difference of where the processing is done. 

We provide examples of using both input delay and processing delay models.
Input delay models are useful for layers where the delay only depends on the
item being delayed.  One example is the delay in \code{CCNxStandardL3}, which is
only a fixed delay.  Another example is \code{CCNxStandardPit}, which uses a
linear function of the CCNxName byte length.
Processing delay models are useful for cases where the delay depends on
the result of the processing.  \code{CCNxStandardFib}, for example, uses
a linear function of the number of FIB lookups done to find the longest matching
prefix.  \code{CCNxStandardContentStore} uses a constant delay if no match is
found and a linear delay in terms of the size of the Content Object matched otherwise.

\begin{figure}
\centering
\includegraphics[width=.95\linewidth, page=13, trim=1.2in 1.6in 0.4in 2.2in, clip]{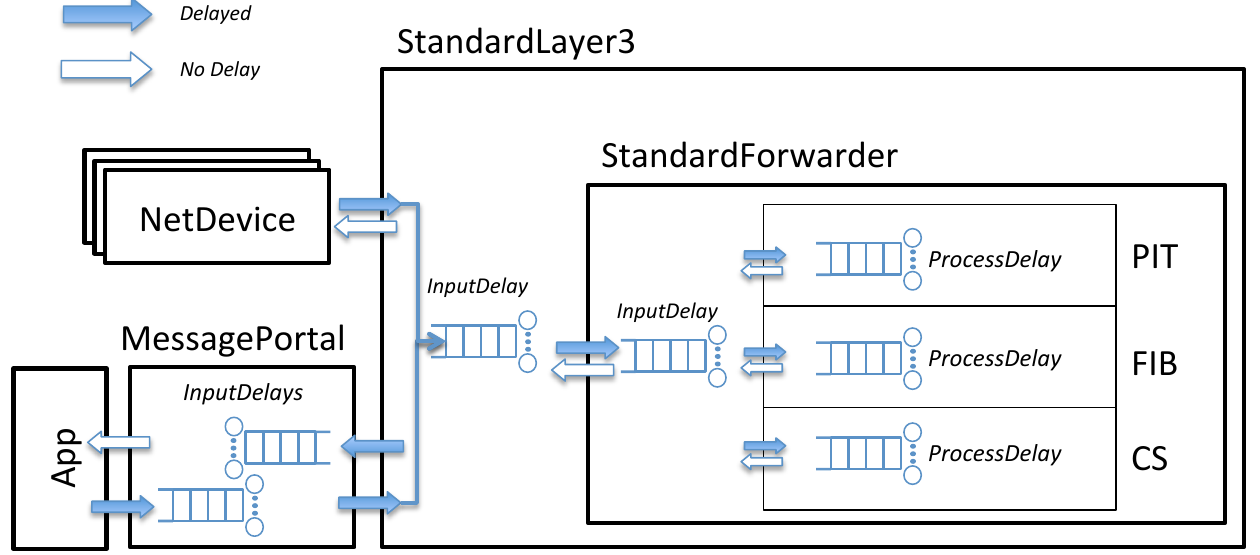}
\caption{Layer delays}
\end{figure}

\begin{figure}
\centering
\includegraphics[width=.95\linewidth, page=10, trim=1.1in 1.5in 0.5in 2.5in, clip]{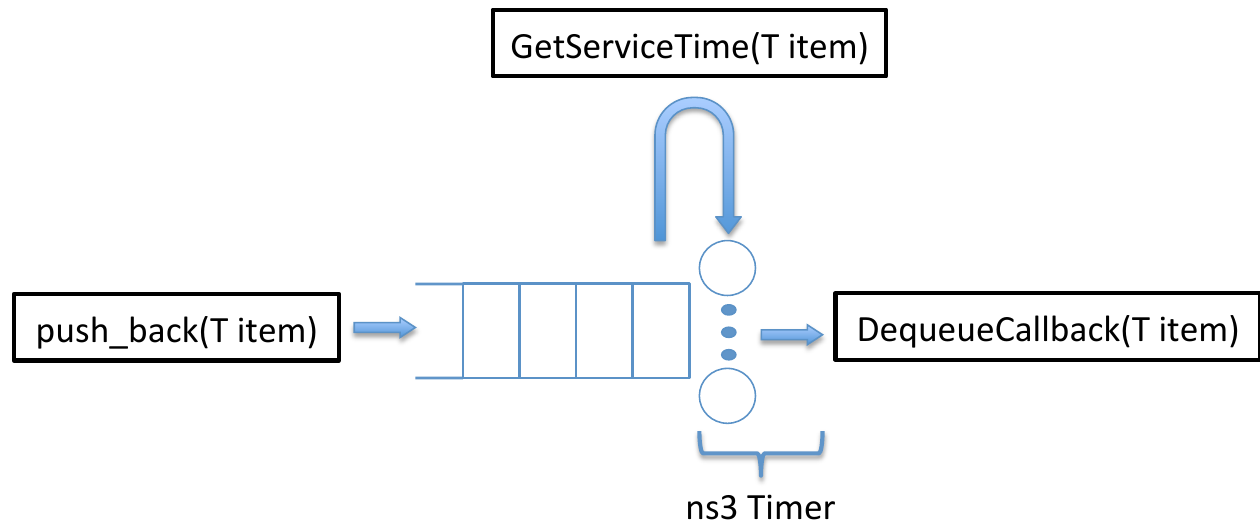}
\caption{Delay Queue}
\end{figure}

A \code{CCNxDelayQueue<T>} is constructed with the number of parallel servers of
the queue (which enforces the delay) and two functions.  The first function
\code{GetServiceTime(T item)} is called whenever a queue item is moved to the head-of-line
and there is a free server for it.  The second function \code{DequeueCallback(T item)}
is called after the service time is over.  It dequeues the item back to the user's code.

\begin{lstlisting}[caption={Modeling input delay},label=code:inputdelay]
Time GetServiceTime(T item) {
  return MicroSeconds(2);
}
void DequeueCallback(T item)  {
  MyResultClass result = lookup(item);
  callNextStage(item, result);
}
\end{lstlisting}
 
When used to model input delay, the \code{GetServiceTime(T item)} function only depends
on the input to the function.  It could be a constant, or some function of the queue
\code{item}.  The delay, for example, could be linear in the number of bytes
of the packet or some function of the number of name components.
An example of a constant delay using this model is in Listing~\ref{code:inputdelay}.
Once the service time is over, the \code{DequeueCallback(T item)} function carries
on processing the queue item.

When used to model processing delay, the delay calculation depends on the
processing function.  A FIB lookup delay, for example, might depend on the
number of name components and the length of each component.
The typical usage pattern is for
\code{GetServiceTime(T item)} to perform the processing on \code{item} and put the result
back in item (e.g. \code{item->SetResult(...)}) and then return the amount of
simulation delay to wait.  Once the delay is over and the item is passed
to \code{DequeueCallback(T item)}, no further calculation is done and \code{item} is
passed to the next step in the service pipeline.  Listing~\ref{code:processingdelay}
shows an example of this usage.
 
 \begin{lstlisting}[caption={Modeling processing delay},label=code:processingdelay]
Time GetServiceTime(T item) {
  MyResultClass result = lookup(item)
  item->SetResult(result);
  return result->GetServiceTime();
}

void DequeueCallback(T item)  {
  callNextStage(item);
}
\end{lstlisting}

\subsection{CCNxL3Protocol (Layer 3)}\label{sec:layer3}
As shown in Figure~\ref{fig:architecture}, the \code{CCNxL3Protocol} interface (abstract base class), implemented by
\code{CCNxStandardlayer3} class, is the central glue between the network and the node.  Besides the necessary APIs to
send and receive packets, it also performs these functions:
\begin{itemize}
\item \emph{Layer 3 Interfaces}: It maintains a layer 3 abstraction of network devices via the \code{CCNxL3Interface} class.  This maintains layer 3 data about an interface, such as if CCNx is forwarding on it or if the interface is administratively up or down.  It also
tracks how to broadcast on a network device.
\item \emph{Connection Table}: The connection table stores information about each adjacency using the \code{CCNxConnection} abstract base class.  It is implemented by \code{CCNxConnectionDevice} for layer 2 devices and \code{CCNxConnectionL4} for layer 4 transport connections.  Each individual peer on a network interface has its own connection.  Each layer 2 interface that supports broadcast has a special connection in the connection table for its broadcast address.
\item \emph{Prefixes}: Registering a prefix from Layer 4 means the FIB will be updated to add to a specific \code{CCNxConnectionL4} for the prefix.  Unregistering a prefix removes that connection from the FIB.
\item \emph{Anchors}: Anchors are like registering and unregistering a prefix, but in addition the routing protocol will be notified that the local node is an anchor.  The routing protocol will then begin advertising the prefix as locally reachable.
\end{itemize}

The \code{CCNxL3Protocol} is replaceable via the simulation script.  A user could create, for example, the \code{AcmeLayer3} that implements the \code{CCNxL3Protocol} interface  and the \code{AcmeLayer3Helper} that implements the \code{CCNxLayer3Helper} abstract base class.  A user may then instantiate the helper in the simulation script and pass it to \code{CCNxStackHelper::SetLayer3Helper()}, as shown in Listing~\ref{code:layer3}.

\begin{lstlisting}[caption={Using a custom Forwarder}, label=code:layer3]
AcmeLayer3Helper layer3;

CCNxStackHelper ccnx;
ccnx.SetLayer3Helper(layer3);
ccnx.Install (nodes);
\end{lstlisting}

\subsection{CCNxStandardForwarder}

The \code{CCNxForwarder} abstract base class represents the interface to a forwarder.  A forwarder manages all the state necessary to find the next hop for  packet.  In a standard implementation, this means it has a FIB to forward Interest and a PIT to forward Content Objects.  The class \code{CCNxStandardForwarder} implements the standards-based forwarder.  The principle function
of the forwarder is to service calls to \code{RouteInput()} and \code{RouteOutput()}.   These functions, whose names are taken from the NS3 Internet module, are the interface to forward a packet.
\code{CCNxL3Protocol} calls the first when it receives a packet from a network device and calls the second when it receives a packet from a Layer 4 protocol.  The result of those functions is a vector of \code{CCNxConnection}s on which forward the packet.

A user may configure the forwarder
via the \code{CCNxStandardForwarderHelper}, which implements the \code{CCNxForwarderHelper} abstract base class.
A user may entirely replace the forwarder by implementing their own forwarder and forwarder helper.  For example, as shown in Listing~\ref{code:forwarder}, the class \code{acme::AcmeForwarder} implements the \code{CCNxForwarder} interface and is
configured via \code{acme::AcmeForwarderHelper}.

\begin{lstlisting}[caption={Using a custom forwarder}, label=code:forwarder]
AcmeForwarderHelper forwarder;
forwarder.UseTwoCopy(true);

CCNxStackHelper ccnx;
ccnx.SetForwardingHelper(forwarder);
ccnx.Install (nodes);
\end{lstlisting}

The \code{CCNxStandardForwarderHelper} exposes three factories to allow a user to customize or replace
the three main functional components: PIT, FIB, and ContentStore.  This abstraction is specific to the \code{CCNxStandardForwarderHelper}, not the \code{CCNxForwarderingHelper} because those tables may not exist
in other forwarders, such as a label-swapping forwarder. 

The tables are represented by an \code{ns3::ObjectFactory}, not a helper.  This is because, in ns3 usage, a helper not only creates an 
\code{ns3::Object}, it also aggregates it to a node.  The factories for the forwarder tables do not perform an aggregation step; 
they are a pure factory, so we call them a factory not a helper.
Suppose you have \code{AcmePit} that implements the \code{CCNxPit} interface.  One may use code similar to Listing~\ref{code:pit} to replace the PIT used by the standard forwarder.

\begin{lstlisting}[caption={Using a custom PIT}, label=code:pit]
AcmePitFactory acmePitFactory;
acmePitFactory.SetMaxEntries(1000);

CCNxStandardForwarderHelper forwarder;
forwarder.SetPitFactory(acmePitFactory);

CCNxStackHelper ccnx;
ccnx.SetForwardingHelper(forwarder);
ccnx.Install (nodes);
\end{lstlisting}

Similar to the PIT, \code{CCNxStandardForwarderHelper} takes an \code{ObjectFactory} for the FIB and Content Store. 
A user may provide their own implementations by creating their own factory and passing it to the 
\code{CCNxStandardForwarderHelper}.

\subsection{CCNxPortal (Layer 4)}\label{sec:layer4}

The \code{ccns3Sim} layer 4 is the \code{CCNxPortal}.  It is modeled after the
\code{ns3::Socket}, which has specializations for UDP and TCP. 
Likewise, Portal has a specialization \code{CCNxMessagePortal}, which is
a simple Interest and Content Object passing portal that does no processing on the
messages.  Future specializations are a Manifest portal and a Chunked portal
that provides reliable, in-order congestion control.

\begin{lstlisting}[caption={Binding a Portal Factory}]
 ObjectFactory f("ns3::ccnx::CCNxMessagePortalFactory");
 Ptr<Object> p = f.Create<Object> ();
 node->AggregateObject (p);
\end{lstlisting}

Similar to \code{ns3::Socket}, when CCNx is instantiated on a node via
\code{CCNxStackHelper}, it creates a \code{CCNxPortalFactory} for
each portal type (in this case just a message portal) and aggregates it to
the node.  This allows applications to ask for the portal factory by name
and create an instance of \code{CCNxPortal} with the desired
implementation.  This method also allows the user to aggregate their
own factories from the simulation script and use those portals in the
same way.

\begin{lstlisting}[caption={Using a Portal},label=code:usingportal]
class Tempor
{
private:
  Ptr<CCNxPortal> m_portal;
  
  void ReceiveNotify(Ptr<CCNxPortal> p) {
    Ptr<CCNxPacket> packet;
    while ((packet = p->Recv ())) {
	  // use packet
    }
  }
public:
  void Labore(Ptr<Node> node) {
    m_portal = CCNxPortal::CreatePortal (node,  TypeId::LookupByName ("ns3::ccnx::CCNxMessagePortalFactory"));

    m_portal->SetRecvCallback (MakeCallback (&Tempor::ReceiveNotify, this));
  }
}; 
\end{lstlisting}

\begin{lstlisting}[caption={CCNxPortal data API}]
// Local forwarder only
virtual bool RegisterPrefix (Ptr<const CCNxName> prefix) = 0;
virtual void UnregisterPrefix (Ptr<const CCNxName> prefix) = 0;

// Local forwarder and routing protocol
virtual bool RegisterAnchor (Ptr<const CCNxName> prefix) = 0;
virtual void UnregisterAnchor (Ptr<const CCNxName> prefix) = 0;

// SendTo() should change to take CCNxConnection
virtual bool Send (Ptr<CCNxPacket> packet) = 0;
virtual bool SendTo (Ptr<CCNxPacket> packet, uint32_t connid) = 0;

virtual Ptr<CCNxPacket> Recv (void) = 0;
virtual Ptr<CCNxPacket> RecvFrom (Ptr<CCNxConnection> & incoming) = 0;
\end{lstlisting}

Listing~\ref{code:usingportal} illustrates a common way to use a
Portal inside a class.  The class defines a private variable \code{m\_portal}
that is instantiated in the method \code{Labore}.  After
the portal is created from its portal factory, the method must also set
the data receive callback, which in this case is \code{Tempor::ReceiveNotify()}.
This design is similar to how the NS3 Socket class works.
When a packet arrives in a portal, it will call the receive notify method
to inform the owner of the portal that data is ready.  The owner of
the portal may begin reading immediately (in the callback) or could
defer reading to a later time.

\section{Routing}\label{sec:routing}

\code{ccns3Sim} provides two ways to do routing.  The first is to use the \code{CCNxStaticRoutingHelper}
to create static routes between two nodes.  The second is to use the NFP routing protocol (Section~\ref{sec:nfp}), which
is  dynamic routing protocol.  A user may also provide their own routing protocol by implementing the
\code{CCNxRoutingProtocol} interface.

A routing protocol uses one or more \code{CCNxPortal}s to send and receive packets from
peers.  It may also use \code{CCNxProtal::SendTo()} to send a packet to a specific connection or rely on the FIB.
A routing protocol uses the \code{CCNxForwarder} interface to call \code{AddRoute()} and \code{RemoveRoute()}
on the forwarder, which allows a very loose coupling between the routing protocol and the forwarder.

A user specifies which routing protocol to use as shown in Listing~\ref{code:routing}.  As with other
components, the user instantiates a routing helper, such as \code{NfpRoutingHelper}, sets any desired
parameters on it, then passes the helper to the \code{CCNxStackHelper}.

\subsection{Installing a routing protocol}
\begin{lstlisting}[caption={Installing a routing protocol}, label=code:routing]
NfpRoutingHelper nfp;
nfp.Set("HelloInterval", TimeValue(Seconds (1)));

CCNxStackHelper ccnx;
ccnx.SetRoutingHelper(forwarder);
ccnx.Install (nodes);
\end{lstlisting}

\subsection{Name Flooding Protocol Routing}\label{sec:nfp}

 The Name Flooding Protocol (NFP) is a distance vector routing protocol for CCNx names.  It creates independent 
 directed acyclic graphs (DAGs) for
 each pair \textit{( anchorName, prefix )}.  An anchor name is a unique identifier for a node that advertises a prefix.  We use
 a CCNxName for both the anchorName and prefix.  The prefix is a CCNx name prefix for matching against Interest messages
 to determine where to send them.  This construction allows each node to keep the maximum information about prefix
 reachability at the expense of extra signaling and storage costs.  Node should still forward only towards the least cost
 anchor, as there is no guarantee that two anchor paths will not cause a loop for the same prefix.  
 NFP is intended as a demonstration of a CCNx routing protocol, and exhibits worse performance than other protocols,
 such as in~\cite{hemmati2015new}.
 
 NFP advertises the tuple \textit{(prefix, anchorName, anchorPrefixSeqnum, distance)}.  Each \textit{(prefix, anchorName)}
  pair is considered independently from
 other advertisements.  This means that NFP maintains routes to all anchors that advertise a specific name prefix.
 Each anchor is responsible for incrementing its own \textit{anchorPrefixSeqnum} with each new advertisement.
 As the advertisement travels over the network, its distance is always increasing with each hop.  We do not
 require min-hop link costs, it could be any positive distance per hop.  
 
  Each anchor has a unique name, for example a cryptographic hash of a public key.
In our experiments, we use a name that is a 32-byte value, similar to what a SHA-256 hash of a public key would be.

 An advertisement for $(prefix, anchorName)$ is considered feasible if the \textit{anchorPrefixSeqnum}is greater than
 any previously seen sequence number, or \textit{anchorPrefixSeqnum} being equal to a previous advertisement and the
 distance is no greater than any previous advertisement seen at the node.  
 An advertisement is \emph{strictly better} if the sequence number is larger or begin equal the distance is strictly less.
 A withdraw message is feasible if its sequence number is larger than the currently known sequence number and it came
 from a current predecessor for the \textit{(prefix, achorName)} pair.  A withdraw has no distance associated with it.
 This feasibility criteria allows equal-cost multipath to
 each anchor.
 Periodically (e.g. every 10 seconds) the anchor advertises its prefixes.
 Each link (neighbor adjacency) has a positive cost.  We use ``1'' for all link costs.

Each routing node will accept only
 feasible advertisements.
 When a node receives a strictly better advertisement, it erases all prior predecessors for the \textit{(prefix, achorName)} pair
 and uses only the strictly better advertisement.  It immediately (with jitter) forwards the advertisement with increased cost
 to its peers.
 When a node receives an equal advertisement, it adds the predecessor to the list of equal cost multi-paths for the 
 \textit{(prefix, achorName)} pair and does not forward the advertisement. 
If a withdraw message changes the state of a \textit{(prefix, achorName)} pair to unreachable, it is forwarded immediately (with jitter).
If it does not change the state, its only effect is to remove a specific predecessor from the set of multi-path predecessors.
 
NFP uses two routing messages: an advertisement and a withdraw.
If a node loses reachability for a route, it may immediately send a withdraw message to its neighbors.  
Routes use soft state and will timeout after 3x the advertisement period.
Advertisements and withdraws are sent in a routing message, which is in turn carried in the payload of a Interest message.
Each routing message has a 1-hop \textit{MessageSeqnum} specific to that 1-hop message.  A message is only accepted if
\textit{(RouterName, MessageSeqnum)} is greater than the previously heard \textit{(RouterName, MessageSeqnum)}.
A routing message is also an implicit Hello from that neighbor.  It may be sent with no advertisements or withdraws
to function simply as a Hello.

The neighbor table tracks Hello status for each neighbor.  It has a callback to \code{NeighborStateChanged()} whenever
the state of a neighbor changes (UP, DOWN, DEAD).  We use a 3-state model for neighbors so we do not erase their
\textit{messageSeqnum} right away when they go DOWN.  They have to timeout a second time to DEAD state before we erase the record.
If a neighbor goes to DOWN state, we erase all next hops that use that neighbor.  That may cause \code{PrefixStateChanged()}
to be called if a \textit{(prefix, anchorName)} becomes unreachable.  On a DEAD change, we simply erase the record.

The prefix table stores the best advertisements we have seen for each \textit{(prefix, anchorName)} pair.
It has a callback to \code{PrefixStateChanged()} when a \textit{(prefix, anchorName)} becomes reachable or unreachable.  There currently is
not a method to prune DEAD records like there is for neighbors.  
 
\section{Applications}\label{sec:apps}
CCNx applications inherit from the \code{CCNxApplication} base class, which in turn inherits from \code{ns3::App\-lication}.
This means that CCNx applications can use the \code{ns3::ApplicationContainer}, which provides easy mechanisms
to start and stop applications on a node.  We provide two example applications: a consumer and a producer.

\subsection{Consumer-Producer}\label{sec:consumerproducer}
The consumer-pro\-duc\-er example applications use a \code{CCNxRepository}.  The repository is a name prefix plus
a fixed number of content objects all of a fixed size.  One repository may be shared between all producers (replicas), so there is no
duplication of state.  Each \code{CCNxProducer} will register as an anchor for the repository prefix (and thus advertise it
via a routing protocol) and serve the repositories content via a \code{CCNxMessagePortal}.  Each \code{CCNxConsumer} will
request names from the repository with a fixed request rate.  For each request, it uses \code{CCNxRepository::GetRandomName()} to
pick a name from the repository.  Currently, the repository only supports a uniform name distribution.
A node may have zero or more producers on it.  

\begin{lstlisting}[caption={Creating a repository}, label=code:repo]
Ptr <const CCNxName> prefix = Create <CCNxName> ("ccnx:/name=prefix");
uint32_t bytes = 124;
uint32_t count = 1000;
Ptr <CCNxContentRepository> repo = Create <CCNxContentRepository> (prefix, bytes, count);
\end{lstlisting}

Listing~\ref{code:repo} shows how to create a content repository with a given prefix.  In this example,
each content object, which will have its own individual name under the name \code{ccnx:/name=prefix}, 
will be 124 bytes of payload, and there will be 1,000 of them.  The \code{Ptr<>} reference to the
repo is then shared between producer nodes and consumer nodes (see the next listing).

Listing~\ref{code:consumer} shows how to create a set of consumer nodes for a repository \code{repo}.
Because the \code{CCNxApplication} class follows the standard NS3 application model, we can use
the \code{ApplicationContainer} to schedule starting and stopping an application.  The same process
applies to the \code{CCNxProducer}. 

\begin{lstlisting}[caption={Configuring a consumer}, label=code:consumer]
CCNxConsumerHelper consumer (repo);
consumer.SetAttribute ("RequestInterval", TimeValue (MilliSeconds (5)));
ApplicationContainer consumerApps = consumer.Install (consumerNodes);
consumerApps.Start (Seconds (2));
consumerApps.Stop (Seconds (12));
\end{lstlisting}

\section{Experiments}\label{sec:experiments}
This section presents results of some experiments with the NFP routing protocol (see Section~\ref{sec:nfp}).
The purpose is to illustrate using the simulator on a non-trivial topology with real routing traffic.
These experiments are not a thorough evaluation of the NFP routing protocol, though they do show some
of its characteristics.  As mentioned in the prior section on NFP, because it treats each pair $( anchorName,
prefix )$ as an independent routing process, it will scale worse than LSCR~\cite{hemmati2015new} or DCR~\cite{Garcia-Luna-Aceves:2014:NCR:2660129.2660141}, which only advertise
the best anchor for a prefix.

We used a similar methodology as in~\cite{hemmati2015new}, which presents the routing protocol LSCR.
The simulations are done on the 154 node AT\&T topology using the simulator's NFP protocol.  
Because NFP is a distance vector protocol, the comparison is not exactly
apples to apples.  Instead of measuring link state advertisements (LSAs), we measure advertisement
and withdraw routing update messages.  We measure computational complexity the same: it is
each call for an event (e.g. a timer or table operation) and each iteration through a loop control structure.
We have not implemented the NLSR or LSCR protocols in our simulator, so we cannot directly
compare results.   The methodology in~\cite{hemmati2015new} uses 30 random nodes as
anchors and randomly distributed 210 name prefixes among them.  We repeat each experiment
20 times.

\begin{figure}
\centering
\includegraphics[width=.95\linewidth, trim=0in 0.0in 0.0in 0.35in, clip]{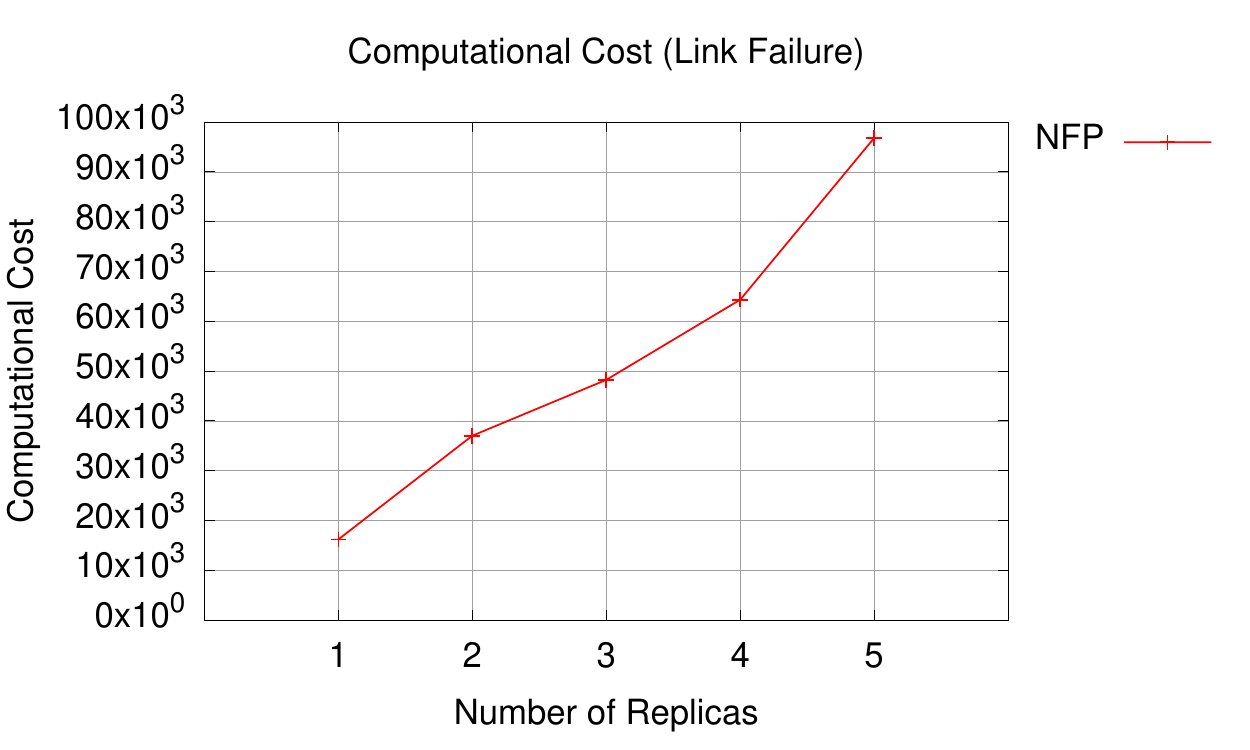}
\caption{Computational Cost after Link Failure}
\label{fig:cost}
\end{figure}

Figure~\ref{fig:cost} shows the computation cost associated with a single link failure.
After an initial period (30 seconds), a single link is selected and failed by introducing
a 100\% packet loss.  After 3 seconds, the Hello protocol times out and the peers
on the link remove each other.  Because the topology has a few nodes of high degree
and many nodes of low degree, this usually results in one or a small number of nodes
being disconnected.  The computational cost is almost linear in the number
of replicas because each $( anchorName, prefix )$ pair is propagated, so when
a leaf node has its link cut, it will lose a number of routes proportional to the number of replicas.

\begin{figure}
\centering
\includegraphics[width=.95\linewidth, trim=0in 0in 0in 0.35in, clip]{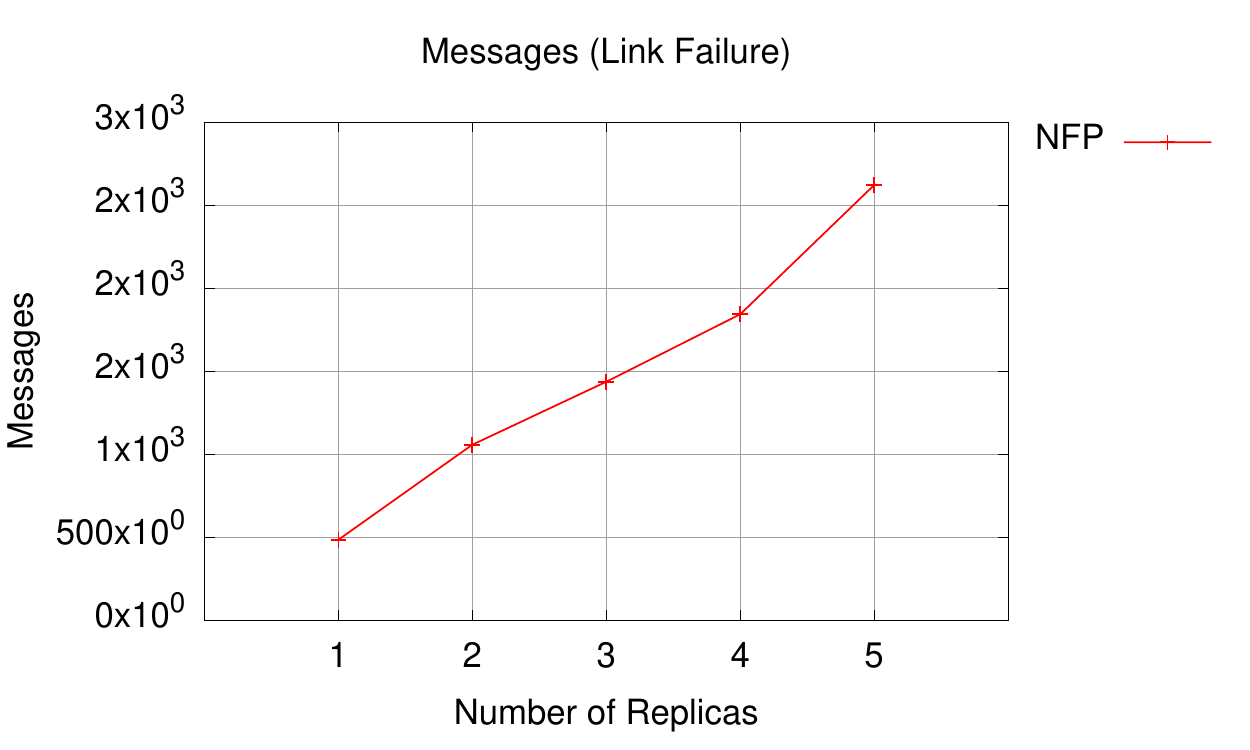}
\caption{Routing messages after Link Failure}
\label{fig:messages}
\end{figure}

Figure~\ref{fig:messages} shows the number of routing update messages triggered
by a link failure.  As with the computational cost, the number of \emph{withdraw} messages
will scale linearly with the number of replicas.

In terms of processing efficiency, \code{ccns3Sim} executes these 50 second virtual time
examples in under 40.1 seconds real time
in a simulation with over 18.8M computational complexity and 221,880 packets.
That comes out to an average execution time of 2.1 usec per routing protocol event
on a MacBook Pro laptop with a 2.6 GHz core i7 CPU.  The simulations use under 285 MB of RAM
(just under 1.8 MB per node).  Most of the memory is going to the content repositories.

\section{Conclusions}\label{sec:conclusion}

The \code{ccns3Sim} module for NS3 implements the CCNx 1.0 protocols. 
It is a native NS3 implementation in C++98 that uses the NS3 coding style,
memory and object management functions, and application style.
Using modular NS3 helpers, a user can substitute or configure all the major
CCNx components: the Layer 3 implementation, the forwarder, the routing protocol,
the PIT, FIB, and Content Store with in the forwarder, and the layer 4 Portal.
Each of these pieces includes a layer delay model to allow simulation
of data structure and processing delays.  We also provide an example
routing protocol, Name Flooding Protocol (NFP) to illustrate how a
dynamic routing protocol operates in \code{ccns3Sim}.

There is still more work to be done.  Future work will
expand the cryptographic side of the implementation to
allow modeling the use of pubic key cryptography.  The packet encoding
currently does not support virtual payloads, which are a common simulation
tool to reduce processing time.  \code{ccns3Sim} is also missing
Interest NACKs, organizational unit TLVs, and nameless objects.
We also plan on incorporating the CCNx model within NS3's LTE model
to simulate its use in mobile networks.

In terms of protocol implementation, future work will include several
important features.  The con\-sum\-er-pro\-duc\-er applications will support
Zipf request distributions and the content store will support different
placement (where to cache) and replacement strategies.  There are also
many optimizations to make content caching simulations more efficient.
We will
also implement several different routing protocols and new transport
protocols, including a chunk based reliable transport and a manifest
based reliable transport.

%
\bibliographystyle{abbrv}
\bibliography{ccns3sim}

\begin{thebibliography}{1}

\bibitem{ccns3Sim}
{ccns3Sim} open source project.
\newblock \url{https://github.com/PARC/ccns3Sim/wiki}.
\newblock Accessed: 2016-05-05.

\bibitem{ns3}
{NS3} simulator.
\newblock \url{http://nsnam.org}.
\newblock Accessed: 2016-05-05.

\bibitem{6654874}
R.~Chiocchetti, D.~Rossi, and G.~Rossini.
\newblock ccnsim: An highly scalable ccn simulator.
\newblock In {\em 2013 IEEE International Conference on Communications (ICC)},
  pages 2309--2314, June 2013.

\bibitem{Garcia-Luna-Aceves:2014:NCR:2660129.2660141}
J.~Garcia-Luna-Aceves.
\newblock Name-based content routing in information centric networks using
  distance information.
\newblock In {\em Proceedings of the 1st International Conference on
  Information-centric Networking}, ICN '14, pages 7--16, New York, NY, USA,
  2014. ACM.

\bibitem{hemmati2015new}
E.~Hemmati and J.~Garcia-Luna-Aceves.
\newblock A new approach to name-based link-state routing for
  information-centric networks.
\newblock In {\em Proceedings of the 2nd International Conference on
  Information-Centric Networking}, pages 29--38. ACM, 2015.

\bibitem{mastorakis2015ndnsim}
S.~Mastorakis, A.~Afanasyev, I.~Moiseenko, and L.~Zhang.
\newblock ndnsim 2.0: A new version of the ndn simulator for ns-3.
\newblock Technical report, Technical Report NDN-0028, NDN, 2015.

\bibitem{I-D.irtf-icnrg-ccnxmessages}
M.~Mosko, I.~Solis, and C.~Wood.
\newblock Ccnx messages in tlv format.
\newblock Internet-Draft draft-irtf-icnrg-ccnxmessages-02, IETF Secretariat,
  April 2016.
\newblock
  \url{http://www.ietf.org/internet-drafts/draft-irtf-icnrg-ccnxmessages-02.txt}.

\bibitem{I-D.irtf-icnrg-ccnxsemantics}
M.~Mosko, I.~Solis, and C.~Wood.
\newblock Ccnx semantics.
\newblock Internet-Draft draft-irtf-icnrg-ccnxsemantics-02, IETF Secretariat,
  April 2016.
\newblock
  \url{http://www.ietf.org/internet-drafts/draft-irtf-icnrg-ccnxsemantics-02.txt}.

\end{thebibliography}

\end{document}